\documentclass{article}
\usepackage{amsfonts}
\usepackage{authblk}
\usepackage{graphicx}
\usepackage{epstopdf}

\begin{document}

\title{Classification of the Lie and Noether point symmetries for the Wave
and the Klein-Gordon equations in pp-wave spacetimes}
\author[1]{A. Paliathanasis\thanks{%
anpaliat@phys.uoa.gr}}
\author[2]{M. Tsamparlis \thanks{%
mtsampa@phys.uoa.gr}}
\author[3]{M.T. Mustafa\thanks{%
tahir.mustafa@qu.edu.qa}}

\affil[1]{Instituto de Ciencias F\'{\i}sicas y Matem\'{a}ticas, Universidad Austral de Chile, Valdivia, Chile}
\affil[2]{Faculty of Physics, Department of Astrophysics-Astronomy-Mechanics, University of Athens, Panepistemiopolis, Athens 157 83, Greece}
\affil[3]{Department of Mathematics, Statistics and Physics, College of Arts and Sciences, Qatar University, Doha 2713, Qatar}

\renewcommand\Authands{ and }
\maketitle

\begin{abstract}
We perform a classification of the Lie and Noether point symmetries for the
Klein-Gordon and for the wave equation in pp-wave spacetimes. To perform
this analysis we reduce the problem of the determination of the point
symmetries to the problem of existence of conformal killing vectors on the
pp-wave spacetimes. We use the existing results of the literature for the
isometry classes of the pp-wave spacetimes and we determine in each class
the functional form of the potential in which the Klein-Gordon equation
admits point symmetries and Noetherian conservation law. Finally we derive
the point symmetries of the wave equation and we find that the maximum
Noether algebra has dimension seven, that is the case of plane wave
spacetimes.\newline
\end{abstract}

\section{Introduction}

Lie point symmetries have been used in order to solve explicitly the
Einstein field equations, and to find new exact solutions in modified
theories of gravity (for instance see \cite%
{LK,Gov,Cap,Dimakis,Yusuf,Maharaj1,Maharaj3} and references therein).
Furthermore, Lie point symmetries have also been used for the study of the
geodesic equations and for the determination of exact solutions of the wave
equation in various gravitation models \cite{FerozeT,GRG2,Camci,Bokh1,Azad}.
\ As far as concerns the wave equation in Riemannian spacetimes, a symmetry
analysis of wave equation in a power-law Bianchi III spacetime can be found
in \cite{Bokhari} and a symmetry analysis of the wave equation on static
spherically symmetric spacetimes, with higher symmetries, was carried out in
\cite{Tahir1}. Recently, in \cite{IJGMMP2}, we started a research program
where we performed the symmetry classification of the wave and the
Klein-Gordon equation in Bianchi I spacetimes. In this work we would like to
extend this analysis and perform a classification of the Lie and Noether
point symmetries for the wave equation and the Klein-Gordon equation in
pp-wave type N spacetimes.

The line element of a pp-wave spacetime is ($A,B=1,2)$ \cite{StephaniBook1}%
\begin{equation}
ds^{2}=-2dudv-2H\left( u,x^{A}\right) du^{2}+\delta _{AB}dx^{A}dx^{B}
\label{pp.01}
\end{equation}%
where $x^{A}=\left( y,z\right) $ are the Cartesian coordinates and $\delta
_{AB}=diag\left( 1,1\right) $ is the two dimensional Euclidian metric. The
non-zero connection coefficients of (\ref{pp.01}) are%
\begin{equation}
\Gamma _{uu}^{v}=H_{,u}~,~\Gamma _{uu}^{A}=H_{,A}~,~\Gamma _{Au}^{v}=H_{,A}.
\end{equation}%
The Laplace operator for the spacetime (\ref{pp.01}) is
\begin{equation}
\Delta \Psi \equiv -2\Psi _{,uv}+2H\left( u,x^{A}\right) \Psi _{,vv}+\Delta
_{\delta }\Psi ,
\end{equation}%
where $\Delta _{\delta }~$is the Laplacian of the two dimensional Euclidian
space. It follows that the Klein-Gordon equation in (\ref{pp.01}) has the
following form%
\begin{equation}
-2\Psi _{,uv}+2H\left( u,x^{A}\right) \Psi _{,vv}+\Delta _{\delta }\Psi
+V\left( u,v,x^{A}\right) \Psi =0.  \label{pp.02}
\end{equation}

The main property of a pp-wave spacetime (\ref{pp.01}) is that admits the
null Killing vector field $k=\partial _{v}$. However for special forms of
the function $H\left( u,x^{A}\right) $ (\ref{pp.01}) admits a greater
conformal algebra. The function $H\left( u,x^{A}\right) $ is computed from
the solution of Einstein field equations. For empty spacetime is given by
the equation $\Delta _{\delta }H=0$, which in Cartesian coordinates is%
\[
H_{,yy}+H_{,zz}=0.
\]%
The classification of the Killing algebras of (\ref{pp.01}) has been done in
\cite{ppKV}, whereas in \cite{ppCKV} are given the conformal algebras of (%
\ref{pp.01}). In the following we use the classification of \cite{ppCKV},
which means that the results we find hold also for non empty spacetimes. \
The plan of the paper is as follows.

In Section \ref{collineations} we give basic definitions and properties of
Riemannian collineations and introduce the Lie and the Noether point
symmetries of differential equations. Furthermore, we discuss the relation
between the Lie and Noether symmetries of the Klein-Gordon equation with the
conformal algebra of the underlying Riemannian manifold. In Section \ref%
{isometryclass} we determine the functional form of the potential $V\left(
u,v,x^{A}\right) $ and the function $H\left( u,x^{A}\right) $ of (\ref{pp.01}%
), in order the Klein-Gordon equation (\ref{pp.02}) to admit Lie and Noether
point symmetries. The complete symmetry analysis for wave equation in the
pp-wave spacetime (\ref{pp.01}) is given in Section \ref{wave}. Finally, in
Section \ref{conclusion} we discuss our results and draw our conclusions.

\section{Collineations and symmetries of differential equations}

\label{collineations}

\subsection{Collineations of Riemannian manifolds}

Let a space with coordinates $\{x^{i}\}.$ Consider the one parameter point
transformation%
\begin{equation}
\bar{x}^{i}=x^{i}+\varepsilon \xi ^{i}\left( x^{k}\right) .  \label{go.02}
\end{equation}%
in which $\xi ^{i}\left( x^{k}\right) $ are the components of a vector field
called the infinitesimal generator of (\ref{go.02}).

Let $\Omega $ be a geometric object in $V^{n}$ with transformation law $%
\Omega ^{a^{\prime }}=\Phi ^{a}\left( \Omega ^{k},x^{k},x^{k^{\prime
}}\right) $.$~\ $Under the action of the point transformation $\Omega $
changes to $\Phi \left( \Omega ^{k},x^{k},x^{k^{\prime }}\right) .$ We
define the Lie derivative $\mathcal{L}_{\xi }$ of $\Omega $ with respect to
the vector field $\xi $ as follows \cite{Yano}
\begin{equation}
\mathcal{L}_{\xi }\Omega =\lim_{\varepsilon \rightarrow 0}\frac{1}{%
\varepsilon }\left[ \Phi \left( \Omega ^{k},x^{k},x^{k^{\prime }}\right)
-\Omega \right] .  \label{go.03}
\end{equation}

By definition the Lie derivative of a geometric object depends on its
transformation law. For functions, the transformation law is $F^{\prime
}\left( \bar{x}^{i}\right) =F\left( x^{i}\right) $, hence under the point
transformation (\ref{go.02}) we have%
\[
\bar{F}\left( \bar{x}^{i}\right) =F\left( x^{i}+\varepsilon \xi ^{i}\right)
=F\left( x^{i}\right) +\varepsilon F_{,i}\xi ^{i}+O\left( \varepsilon
^{2}\right) .
\]

Hence from (\ref{go.03}) it follows
\begin{equation}
\mathcal{L}_{\xi }F=F_{,i}\xi ^{i}.  \label{go.04}
\end{equation}%
We say that the function $F\left( x^{i}\right) $ is invariant under the
action of (\ref{go.02}) if $\mathcal{L}_{\xi }F=0$; In this case $\xi $ is
called a symmetry of the function $F\left( x^{i}\right) .$

In general we have%
\begin{equation}
\mathcal{L}_{\xi }\Omega =\Lambda  \label{go.05}
\end{equation}%
where $\Lambda $ is a tensor which has the same number and symmetries of the
indices with $\Omega $. We remark that $\Omega $ is not necessarily a
tensor. In this case we say that the vector field $\xi $ is a collineation
of $\Omega .$ The type of collineations depends on the tensor $\Lambda $.

In Riemannian Geometry (and in General Relativity) in general we are
interested on geometrical objects $\Omega $ which are defined in terms of
the metric\footnote{%
For the complete classification of the collineations of Riemannian manifold
see \cite{Katzin69,GSHall}.}. In particular in this work we shall consider
the geometric object $\Omega =$ $g_{ij},$ and $\Lambda =$ $2\psi \left(
x^{k}\right) g_{ij}$; that is, condition (\ref{go.05}) becomes%
\begin{equation}
\mathcal{L}_{\xi }g_{ij}=2\psi \left( x^{k}\right) g_{ij}.  \label{go.07}
\end{equation}

The vector field $\xi $ is called as conformal Killing vector (CKV)$.$ In
general $\psi \left( x^{k}\right) =\frac{1}{n}\xi _{;i}^{i}$, \ where $n$ is
the dimension of the Riemann space $V^{n}$ and $";"$ denotes the covariant
derivative with respect to the metric tensor $g_{ij}$. If $\psi _{;ij}=0$
the field $\xi $ is called special Conformal Killing vector (sp.CKV), when $%
\psi _{;i}=0$, i.e. $\psi \left( x^{k}\right) =\psi _{0}$, $\xi $ is called
Homothetic vector (HV) and when $\psi \left( x^{k}\right) =0$, $\xi $ is a
Killing vector (KV) of the metric tensor $g_{ij}$.

The CKVs\ of a metric form a Lie algebra and so do the KVs and the HV. If we
denote by $G_{CV},G_{HV}$, $G_{KV}$ these algebras we have the inclusion
relations
\begin{equation}
G_{KV}\subseteq G_{HV}\subseteq G_{CV}  \label{go.08}
\end{equation}%
and~
\begin{equation}
0\leq \dim G_{H-K}\leq 1~  \label{go.09}
\end{equation}%
where $G_{H-K}=G_{HV}-G_{HV}\cap G_{KV};\ $the last relation means that a
Riemannian space admits at most one HV. Concerning the dimension of the
conformal algebra $G_{CV}$ we have $\dim G_{CV}\leq \frac{1}{2}\left(
n+1\right) \left( n+2\right) .$

Collineations constitute a strong constraint on the geometric structure of a
space. For example if a space admits $\frac{1}{2}n\left( n+1\right) $ then
it must be a space of constant curvature and there are only three types of
spaces with curvature $0,\pm K$ whose metric in Cartesian coordinates has
the general form
\begin{equation}
ds^{2}=\left( 1+\frac{K}{4}\eta _{ij}x^{i}x^{j}\right) ^{-1}\left( \eta
_{ij}dx^{i}dx^{j}\right) .  \label{go.09a}
\end{equation}

\subsection{Lie point symmetries of differential equations}

Consider the second order partial differential equation $\Theta \left(
x^{k},\Psi ,\Psi _{,i},\Psi _{,ij}\right) =0$, where $x^{k}$ are the
independent variables, $\Psi =\Psi \left( x^{k}\right) $ is the dependent
variable and, $\Psi _{,i}=\frac{\partial \Psi }{\partial x^{i}}$. Latin
indices take the values $1,2,...,n$. \ Let $\left( x^{i},\Psi \right)
\rightarrow \left( \bar{x}^{i}\left( x^{k},\Psi ,\varepsilon \right) ,\bar{%
\Psi}\left( x^{k},\Psi ,\varepsilon \right) \right) ,$ be a one parameter
point transformation of the independent and dependent variables with
infinitesimal generator
\begin{equation}
X=\xi ^{i}\left( x^{k},\Psi \right) \partial _{i}+\eta \left( x^{k},\Psi
\right) \partial _{\Psi }.  \label{go.10}
\end{equation}

The differential equation $\Theta $ can be seen as a geometric object on the
jet space $J=J\left( x^{k},\Psi ,\Psi _{,i},\Psi _{,ij}\right) .$ We say
that $X$ defines a Lie point symmetry of $\Theta \left( x^{k},\Psi ,\Psi
_{,i},\Psi _{,ij}\right) =0,$ if the following condition is satisfied%
\begin{equation}
\mathcal{L}_{X^{\left[ 2\right] }}\Theta =\lambda \Theta ,  \label{go.12}
\end{equation}%
in which $X^{\left[ 2\right] },$ is the second extension/prolongation of $X$
in the space $J$, given by the formula
\begin{equation}
X^{\left[ 2\right] }=X+\eta _{i}\partial _{\Psi _{,i}}+\eta _{ij}\partial
_{\Psi _{,i}},  \label{go.13}
\end{equation}%
where,$~\eta _{i}=D_{i}\left( \eta \right) -\Psi _{,k}D_{i}\left( \xi
^{k}\right) $, $\eta _{ij}=D_{j}\left( \eta _{i}\right) -\Psi _{ki}D_{j}\xi
^{k}$, and $D_{i}$ is the operator of the total derivative, i.e. $D_{i}=%
\frac{\partial }{\partial x^{i}}+\Psi _{,i}\frac{\partial }{\partial \Psi }%
+\Psi _{,ij}\frac{\partial }{\partial \Psi _{,j}}+...~$\cite{Bluman,Ibrag}.

The existence of a Lie point symmetry for a partial differential equation
(PDE)\ means that there exist a "coordinate" system in which the
differential equation $\Theta $ is independent on one of the independent
variables. In addition, Lie point symmetries can be used in order to
transform solutions into solutions between different points of the space $%
\left( x^{i},\Psi \right) ~$\cite{Bluman3}.

If the differential equation follows form a Lagrangian $L=L\left( x^{k},\Psi
,\Psi _{,i}\right) $, that is $\Theta \equiv \mathbf{E}\left( L\right) =0$,
where $\mathbf{E}$ is the Euler-operator, then one defines a special type of
Lie symmetry by the condition
\begin{equation}
\mathcal{L}_{X^{\left[ 1\right] }}L+LD_{i}\xi ^{i}=D_{i}A^{i},  \label{go.14}
\end{equation}%
where $A^{i}$ is a vector field and $X^{\left[ 1\right] }$ \ is the first
prolongation of $X$. These Lie point symmetries are called Noether point
symmetries\footnote{%
In the literature the vector fields which satisfy condition (\ref{go.14})
have been called Noether Gauge symmetries. However, condition (\ref{go.14})
is that of the standard Noether's theorem \cite{Emmy} and the use of the
term Gauge is unnecessar, for instance see \cite{NG01,NG02,NG03}.} and have
the characteristic property that to the Lie symmetry $X$ there corresponds a
conserved current $I^{i}\left( x^{k},\Psi ,\Psi _{,i}\right) $, that is $%
D_{i}I^{i}=0$, where
\begin{equation}
I_{i}=\xi ^{j}\mathcal{H}_{ij}+\eta p_{i}-A_{i}.  \label{go.15}
\end{equation}%
where $p^{i}=\frac{\partial L}{\partial \Psi _{,i}}$ and $\mathcal{H}%
_{~j}^{i}=\frac{\partial L}{\partial \Psi _{,i}}\Psi _{,j}-L$.

If (\ref{go.14}) holds, the Lie symmetry $X$ is called Noether symmetry and
the vector field $I^{i}$ Noether current. The Lie point symmetries of a PDE\
form a Lie algebra and the Noether point symmetries a subalgebra of this
algebra.

\subsection{Collineations of Riemannian spaces as point symmetries of the
Klein-Gordon equation}

One parameter point transformations in a Riemannian space define the
collineations in that space which \ characterize to a large extent the
geometry of the space. But one parameter point transformations define also
the Lie point symmetries of PDEs in that space. Therefore one should expect
that there exists a relation between the collineations of a space and the
Lie / Noether point symmetries of a PDE\ in that space. The reason for this
is twofold. First one may "see " the defining equation of a collineation as
a PDE\ in the space which remains invariant under the Lie point symmetry.
Second the dynamical field equations which describe the evolution of a
dynamical system the space should be affected by the geometry of the space.
This is most vividly seen in the case of the geodesic equations which on one
hand characterize the geometry of the space and on the other, in accordance
to the Principle of Equivalence, describe the equations of motion of a
"free" particle in the space.

Indeed it has been shown that Lie point symmetries of the geodesic equations
in a Riemannian space are elements of the special Projective algebra of the
space \cite{GRG2}, and that the Lie point symmetries form the projective
algebra of an extended manifold, for details see \cite{Aminova1,Aminova2}.

Similar results have been found for partial differential equations which
involve the metric tensor $g_{ij}$. \ Specifically, it has been shown that
the Lie point symmetries of the Schr\"{o}dinger equation are generated by
the Homothetic algebra of the space which defines the Laplace operator \cite%
{IJGMMP}. Moreover, the Lie point symmetries of the wave and of the Poisson
equation are elements of the conformal algebra of the Riemannian manifold
\cite{Ibrag,Bozhkov}.

The Klein-Gordon equation in a general Riemannian space is defined as
follows
\begin{equation}
\Delta _{g}\Psi +V\left( x^{k}\right) \Psi =0,  \label{kg.01}
\end{equation}%
where $\Delta _{g}=\frac{1}{\sqrt{\left\vert g\right\vert }}\frac{\partial }{%
\partial x^{i}}\left( \sqrt{\left\vert g\right\vert }g^{ij}\frac{\partial }{%
\partial x^{j}}\right) $, is the Laplace operator defined by the metric
tensor $g_{ij}$. Equation (\ref{kg.01}) arise from a variational principle
given by the following Lagrangian function%
\begin{equation}
L\left( x^{k},\Psi ,\Psi _{,k}\right) =\frac{1}{2}\sqrt{\left\vert
g\right\vert }g^{ij}\Psi _{,i}\Psi _{,j}-\frac{1}{2}\sqrt{\left\vert
g\right\vert }V\left( x^{k}\right) \Psi ^{2}  \label{kg.02}
\end{equation}

In \cite{IJGMMP}, it has been shown that the Lie point symmetries for the
Klein-Gordon equation (\ref{kg.01}) are%
\begin{equation}
X=\xi ^{i}\left( x^{k}\right) \partial _{i}+\left( \frac{2-n}{2}\psi \left(
x^{k}\right) \Psi \right) \partial _{\Psi }~,~X_{\Psi }=\Psi \partial _{\Psi
}~,~X_{b}=b\left( x^{k}\right) \partial _{\Psi }  \label{kg.03}
\end{equation}%
where $\xi ^{i}\left( x^{k}\right) $ is a CKV of $g_{ij}$ with conformal
factor $\psi \left( x^{k}\right) ,$ $b\left( x^{k}\right) $ is a solution of
the original equation (\ref{kg.01}) and the following condition holds
\begin{equation}
\xi ^{k}V_{,k}+2\psi V-\frac{2-n}{2}\Delta _{g}\psi =0  \label{kg.04}
\end{equation}%
where $n=\dim g_{ij}$. The two fields $X_{\Psi },X_{b}$ are called linear
symmetries, because they exist for a general linear partial equation. For
obvious reasons $X_{b}$ is called a solution symmetry.

Concerning the Noether point symmetries of (\ref{kg.01}) we have that for
the Lagrange function (\ref{kg.02}), the Lie point symmetries of (\ref{kg.01}%
) (except the two trivial ones) are also Noether point symmetries of (\ref%
{kg.02}) and that the field $A^{i}$ of condition (\ref{go.14}) has the
following form
\begin{equation}
A_{i}=\frac{2-n}{4}\sqrt{g}\psi _{,i}\left( x^{k}\right) u^{2}.
\label{kg.05}
\end{equation}

In the following we apply these results in order to classify the Lie and the
Noether point symmetries of the Klein-Gordon equation (\ref{kg.01}) and the
wave equation, $V\left( x^{k}\right) =0$, in pp-wave spacetimes.

\section{Lie and Noether point symmetries of the Klein-Gordon equation in
pp-wave spacetimes}

\label{isometryclass}

The pp-wave spacetimes have been classified according to the admitted
isometry algebra in \cite{ppKV}. The complete classification of the CKVs for
the pp-wave spacetimes has been done in \cite{ppCKV,ppnull}. However, as we
discussed above the Lie/Noether point symmetries of the Klein-Gordon
equation follow from the conformal algebra of the space which defines the
Laplace operator which means that in order to perform the classification
problem we follow the results of \cite{ppCKV,ppnull} in order to determine
all potentials $V\left( u,v,x^{k}\right) $ for which the resulting
Klein-Gordon equation (\ref{pp.02}) admits Lie and Noether point symmetries.

From \cite{ppKV} and \cite{ppCKV}, we have 14 isometry classes with some
subclasses, in which the spacetime (\ref{pp.01}) admits a greater conformal
algebra. We remark that equation (\ref{pp.02}) is a linear equation
therefore admits always the linear (trivial) symmetries $X_{\Psi },~X_{b}$.

\subsection{Isometry class 1}

This is the most general isometry class and $H=H\left( u,x^{A}\right) $ is
an arbitrary function. The spacetime (\ref{pp.01}) admits a one dimensional
conformal algebra given by the KV $k$. Hence we have that the Klein-Gordon
equation (\ref{pp.02}) admits the vector field $k$ as a Lie or Noether point
symmetry if and only if,
\begin{equation}
\mathcal{L}_{k}V=0,
\end{equation}%
from which follows that $V_{G}^{\left( 1\right) }=V\left( u,x,y\right) $.

\subsubsection{Subclass 1i}

If, $H=H\left( u,z\right) ,$ space (\ref{pp.01}) admits a three dimensional
conformal algebra spanned by the three KVs
\begin{equation}
k=\partial_{v}~,~X_{2}^{\left( 1i\right) }=\partial_{y}~,~X_{3}^{\left(
1i\right) }=y\partial _{v}+u\partial _{y}.
\end{equation}%
Therefore from conditions (\ref{kg.04}) we have that $X_{2}^{\left(
1i\right) },$ $X_{3}^{\left( 1i\right) }$ are Lie point symmetries when%
\begin{eqnarray}
X_{2}^{\left( 1i\right) } &:&V_{2}^{\left( 1i\right) }=V\left( u,v,z\right) ,
\\
X_{3}^{\left( 1i\right) } &:&V_{3}^{\left( 1i\right) }=V\left(
u,x,y^{2}-2uv\right) .
\end{eqnarray}

Furthermore from the linear combinations of the vector fields we have that
the vector $X_{\left( 1i\right) }=c_{1}k+c_{2}X_{2}^{\left( 1i\right)
}+c_{3}X_{3}^{\left( 1i\right) }$ is a Lie point symmetry of (\ref{pp.02})
when%
\begin{equation}
V_{G}^{\left( 1i\right) }=\left( u,x,\frac{2c_{1}y-2c_{2}v+c_{3}\left(
y^{2}-2uv\right) }{2\left( c_{2}+c_{3}u\right) }\right) .
\end{equation}%
The commutators of the Killing algebra are as follows%
\begin{equation}
\left[ k,X_{2}^{\left( 1i\right) }\right] =0~,~\left[ k,X_{3}^{\left(
1i\right) }\right] =0~,~\left[ X_{2}^{\left( 1i\right) },X_{3}^{\left(
1i\right) }\right] =k
\end{equation}

\subsection{Isometry class 2}

When $H=H\left( u,r\right) ,$ the space admits a two dimensional conformal
algebra spanned by the KVs $k$ and $X_{2}^{\left( 2\right) }=\partial
_{\theta }$. Therefore we have that the field $X_{\left( 2\right)
}=c_{1}k+c_{2}X_{2}^{\left( 2\right) }$ is a Lie point symmetry of (\ref%
{pp.02}), if and only if,
\begin{equation}
V_{2}^{\left( 2\right) }=V\left( u,r,\theta -\frac{c_{2}}{c_{1}}v\right) .
\end{equation}

In this isometry class correspond four subclasses where the space (\ref%
{pp.01}) admits a greater conformal algebra. However in the fourth class the
exact form of the function $H\left( u,r\right) $ is not exact, hence we
study only the three subclasses.

\subsubsection{Subclass 2i}

Assume that $H=K\left( \alpha u+\beta \right) ^{q}\ln r,$ with $\beta \in
\mathbb{R}
$ and $K,\alpha \in
\mathbb{R}
^{\ast }$. In this case the space (\ref{pp.01}) admits an extra HV.

For $q\neq -1$ the HV vector is
\begin{equation}
H_{3}^{\left( 2i\right) }=\frac{2}{2\alpha +\alpha q}\left[ \left( \alpha
u+\beta \right) \partial _{u}+\left( \alpha \left( q+q\right) v-\frac{q+2}{%
2\left( q+1\right) }K\left( \alpha u+\beta \right) ^{q+1}\right) \partial
_{v}+\frac{2+q}{2}\alpha r\partial _{r}\right]
\end{equation}%
whereas for $q=-1$ the HV is%
\begin{equation}
H_{3\left( -1\right) }^{\left( 2i\right) }=\frac{2}{\alpha }\left[ \left(
\alpha u+\beta \right) \partial _{u}-\frac{K}{2}\ln \left( \alpha u+\beta
\right) \partial _{v}+\alpha r\partial _{r}\right]
\end{equation}%
The corresponding factors $\psi _{H}^{\left( 2i\right) }~$and~$\psi
_{H}^{\left( 2i-1\right) }$ are equal to one.

We have that the vector fields $Y_{3}^{\left( 2i\right) }=H_{3}^{\left(
2i\right) }-\frac{1}{2}X_{\Psi }$, $Y_{3\left( -1\right) }^{\left( 2i\right)
}=H_{3\left( -1\right) }^{\left( 2i\right) }-\frac{1}{2}X_{\Psi }$ are Lie
point symmetries of (\ref{pp.02}) provided the potential has the form%
\begin{equation}
Y_{3}^{\left( 2i\right) }:V_{3}^{\left( 2i\right) }=\left( \alpha u+\beta
\right) ^{-\left( q+2\right) }V\left( v\left( \alpha u+\beta \right)
^{-\left( q+1\right) }+\frac{q+2}{2\alpha \left( q+1\right) }K\ln \left(
\alpha u+\beta \right) ,r\left( \alpha u+\beta \right) ^{-1-\frac{q}{2}%
},\theta \right) ,
\end{equation}%
and%
\begin{equation}
Y_{3\left( -1\right) }^{\left( 2i\right) }:V_{3\left( -1\right) }^{\left(
2i\right) }=\left( \alpha u+\beta \right) ^{-1}V\left( v+\frac{K}{4\alpha }%
\left( \ln \left( \alpha u+\beta \right) \right) ^{2},\frac{r}{\sqrt{\alpha
u+\beta }},\theta \right) .
\end{equation}

The generic fields $X_{\left( 2i\right) }=c_{1}k+c_{2}X_{2}^{\left( 2\right)
}+c_{3}H^{\left( 2i\right) }-\frac{1}{2}c_{3}X_{\Psi }$ and $X_{\left(
2i-1\right) }$\thinspace $=c_{1}k+c_{2}X_{2}^{\left( 2\right)
}+c_{3}H_{\left( -1\right) }^{\left( 2i\right) }-\frac{1}{2}c_{3}X_{\Psi }$
are Lie point symmetries when the potential is
\begin{equation}
X_{\left( 2i\right) }:V_{G}^{\left( 2i\right) }=\left( \alpha u+\beta
\right) ^{-\left( q+2\right) }V\left( f\left( u,v\right) ,r\left( \alpha
u+\beta \right) ^{-1-\frac{q}{2}},\theta -\frac{c_{2}\left( q+2\right) }{%
2c_{3}}\ln \left( \alpha u+\beta \right) \right) ,
\end{equation}%
or%
\begin{equation}
X_{\left( 2i-1\right) }:V_{G\left( -1\right) }^{\left( 2i\right) }=\left(
\alpha u+\beta \right) ^{-1}V\left( v-\frac{c_{1}}{2c_{3}}\ln \left( \alpha
u+\beta \right) +\frac{K}{4\alpha }\left( \ln \left( \alpha u+\beta \right)
\right) ^{2},\frac{r}{\sqrt{\alpha u+\beta }},\theta -\frac{c_{2}}{2c_{3}}%
\ln \left( \alpha u+\beta \right) \right) ,
\end{equation}%
respectively; the function $f\left( u,v\right) $ is
\begin{equation}
f\left( u,v\right) =\left( v+\frac{c_{1}\left( q+2\right) }{2c_{3}\left(
q+1\right) }\right) \left( \alpha u+\beta \right) ^{-\left( q+1\right) }+%
\frac{q+2}{2\alpha \left( q+1\right) }K\ln \left( \alpha u+\beta \right) .
\end{equation}

\subsubsection{Subclass 2ii}

When $H=Ke^{-\frac{\Theta }{\beta }u}\ln r$, where$~\Theta \in
\mathbb{R}
$ and $\beta ,K\in
\mathbb{R}
^{\ast },$ the space (\ref{pp.01}) admits an extra HV. For $\Theta \neq 0$
the HV is%
\begin{equation}
H_{3}^{\left( 2ii\right) }=-\frac{2\beta }{\Theta }\partial _{u}+\left( 2v+%
\frac{\beta }{\Theta }Ke^{-\frac{\Theta }{\beta }u}\right) \partial
_{v}+r\partial _{r}~,~\psi _{H}^{\left( 2ii\right) }=1,
\end{equation}%
and for $\Theta =0$ the HV is%
\begin{equation}
H_{3\left( 0\right) }^{\left( 2ii\right) }=u\partial _{u}+\left( v-Ku\right)
\partial _{v}+r\partial _{r}~,~\psi _{H}^{\left( 2ii0\right) }=1.
\end{equation}

Hence, we have that $Y_{3}^{\left( 2ii\right) }=H^{\left( 2ii\right) }-\frac{%
1}{2}X_{\Psi },~Y_{3\left( 0\right) }^{\left( 2ii\right) }=H_{\left(
0\right) }^{\left( 2ii\right) }-\frac{1}{2}X_{\Psi }$ are Lie point
symmetries of (\ref{pp.02}) if and only if the potential is

\begin{eqnarray}
Y_{3}^{\left( 2ii\right) } &:&V_{3}^{\left( 2ii\right) }=V\left( ve^{\frac{%
\Theta }{\beta }u}+\frac{K}{2}u,re^{\frac{\Theta }{\beta }u},\theta \right)
e^{\frac{\Theta }{\beta }u}, \\
Y_{3\left( 0\right) }^{\left( 2ii\right) } &:&V_{3\left( 0\right) }^{\left(
2ii\right) }=V\left( vu^{-1}+K\ln u,ru^{-1},\theta \right) .
\end{eqnarray}%
The generic vector fields $X_{\left( 2ii\right) }=c_{1}k+c_{2}X_{2}^{\left(
2\right) }+c_{3}Y_{3}^{\left( 2ii\right) }$ and $X_{\left( 2ii0\right)
}=c_{1}k+c_{2}X_{2}^{\left( 2\right) }+c_{3}Y_{3\left( 0\right) }^{\left(
2ii\right) },$ are Lie point symmetries of the Klein Gordon equation if
\begin{eqnarray}
X_{\left( 2ii\right) } &:&V_{G}^{\left( 2ii\right) }=V\left( \left( v+\frac{%
c_{1}}{2c_{3}}\right) e^{\frac{\Theta }{\beta }u}+\frac{K}{2}u,re^{\frac{%
\Theta }{\beta }u},\theta -\frac{c_{2}}{2c_{3}}\frac{\Theta }{\beta }%
v\right) e^{\frac{\Theta }{\beta }u}, \\
X_{\left( 2ii0\right) } &:&V_{G}^{\left( 2ii0\right) }=u^{-2}V\left( \left(
v+\frac{c_{1}}{c_{3}}\right) u^{-1}+K\ln u,ru^{-1},\theta -\frac{c_{2}}{c_{3}%
}\ln u\right) .
\end{eqnarray}

\subsubsection{Subclass 2iii}

When $H=e^{g\left( u\right) }\ln r$ with $g\left( u\right) =-\ln \left( \rho
u^{2}+\alpha u+\beta \right) -\Theta \int \left( \rho u^{2}+\alpha u+\beta
\right) ^{-1}du$ and $\rho ,\alpha ,\beta \in
\mathbb{R}
^{\ast },~\Theta \in
\mathbb{R}
$ the space (\ref{pp.01}) admits a sp.CKV. For simplicity we study the case $%
\Theta =0$. The sp.CKV is
\begin{equation}
S_{3}^{\left( 2iii\right) }=2e^{-g\left( u\right) }\partial _{u}+\left( \rho
r^{2}+g\left( u\right) \right) \partial _{v}+\left( 2\rho u+\alpha \right)
r\partial _{r},
\end{equation}%
where the conformal factor is $\psi _{S}^{\left( 2iii\right) }=2\rho
u+\alpha ,$ with $\left( \psi _{S}^{\left( 2iii\right) }\right) _{;\mu \nu
}=0$.

Therefore from the sp.CKV $S_{3}^{\left( 2iii\right) },$ the generated
Lie/Noether point symmetry vector is $Y_{3}^{\left( 2iii\right)
}=S_{3}^{\left( 2iii\right) }-\frac{1}{2}\psi _{S}^{\left( 2iii\right)
}X_{\Psi }$ with corresponding potential%
\begin{equation}
V_{3}^{\left( 2iii\right) }=e^{g\left( u\right) }V\left( v-\frac{\rho }{2}%
r^{2}ue^{g\left( u\right) }-\int g\left( u\right) e^{g\left( u\right)
}du,re^{\frac{1}{2}g\left( u\right) },\theta \right) .
\end{equation}

Furthermore the vector field $Y_{\left( 2iii\right) }=c_{1}k+c_{2}X_{\left(
2\right) }^{2}+c_{3}Y_{3}^{\left( 2iii\right) }$ is a Lie point symmetry
vector of (\ref{pp.02}) if%
\begin{equation}
V_{G}^{\left( 2iii\right) }=e^{-c_{3}g\left( u\right) }V\left( v-\frac{\rho
}{2}r^{2}ue^{g\left( u\right) }-\int e^{g\left( u\right) }\left( g\left(
u\right) +\frac{c_{1}}{c_{3}}\right) du,~re^{\frac{1}{2}g\left( u\right)
},~\theta -\frac{c_{2}}{c_{3}}\frac{\arctan \left( \frac{\psi _{S}^{\left(
2iii\right) }}{\sqrt{4\rho \beta -\alpha ^{2}}}\right) }{\sqrt{4\rho \beta
-\alpha ^{2}}}\right) .
\end{equation}

The commutators of the elements of the conformal algebras of the isometry
class 2 with the subclasses (2i)-(2iii) are given in table \ref{comClass2}.

\begin{table}[tbp] \centering%
\caption{Commutators of the conformal algebras of isometry class 2 of the
pp-wave spacetime (\ref{pp.01}) and of the subclasses (2i)-(2iii)}%
\begin{tabular}{c|ccccccc}
\hline\hline
$\left[ \mathbf{X}_{I},\mathbf{X}_{J}\right] $ & $k$ & $X_{2}^{\left(
2\right) }$ & $H_{3}^{\left( 2i\right) }$ & $H_{3\left( -1\right) }^{\left(
2i\right) }$ & $H_{3}^{\left( 2ii\right) }$ & $H_{3\left( 0\right) }^{\left(
2ii\right) }$ & $S_{3}^{\left( 2iii\right) }$ \\ \hline
$k$ & $0$ & $0$ & $2\frac{q+1}{q+2}k$ & $0$ & $2k$ & $k$ & $0$ \\
$X_{2}^{\left( 2\right) }$ & $0$ & $0$ & $0$ & $0$ & $0$ & $0$ & $0$ \\
$H_{3}^{\left( 2i\right) }$ & $-2\frac{q+1}{q+2}k$ & $0$ & $0$ &  &  &  &
\\
$H_{3\left( -1\right) }^{\left( 2i\right) }$ & $0$ & $0$ &  & $0$ &  &  &
\\
$H_{3}^{\left( 2ii\right) }$ & $-2k$ & $0$ &  &  & $0$ &  &  \\
$H_{3\left( 0\right) }^{\left( 2ii\right) }$ & $-k$ & $0$ &  &  &  & $0$ &
\\
$S_{3}^{\left( 2iii\right) }$ & $0$ & $0$ &  &  &  &  & $0$ \\ \hline\hline
\end{tabular}%
\label{comClass2}%
\end{table}%

\subsection{Isometry class 3}

In the isometry class 3, the function $H\left( u,x^{A}\right) $ is of the
form $H=u^{-2}W\left( s,t\right) $ where the coordinates $\left\{
s,t\right\} $ are
\begin{eqnarray}
s &=&y\sin \left( c\ln u\right) -z\cos \left( c\ln u\right) , \\
~t &=&y\cos \left( c\ln u\right) +z\sin \left( c\ln u\right) .
\end{eqnarray}

The pp-wave spacetime (\ref{pp.01}) admits two KVs, the fields $k$ and $%
X_{2}^{\left( 3\right) }=u\partial _{u}-v\partial _{v}$ with commutator $%
\left[ k,X_{2}^{\left( 3\right) }\right] =-k$. \ Hence the vector field $%
X_{3}=c_{1}k+c_{2}X_{2}^{\left( 3\right) }$ is a Lie point symmetry of (\ref%
{pp.02}) provided that
\begin{equation}
V_{G}^{\left( 3\right) }=V\left( uv-\frac{c_{1}}{c_{2}}u,s,t\right) .
\end{equation}

\subsection{Isometry class 4}

When $H=W\left( \bar{s},\bar{t}\right) $ with%
\begin{eqnarray}
\bar{s} &=&y\sin \left( cu\right) -z\cos \left( cu\right) \\
\bar{t} &=&y\cos \left( cu\right) +z\sin \left( cu\right)
\end{eqnarray}%
the pp-wave spacetime (\ref{pp.01}) admits a two dimensional conformal
algebra with elements the two KVs $k,~X_{2}^{\left( 4\right) }=\partial _{u}$
with commutator $\left[ k,X_{2}^{\left( 4\right) }\right] =0$. Therefore we
have that the field $X_{4}=c_{1}k+c_{2}X_{2}^{\left( 4\right) }$ is a Lie
point symmetry of (\ref{pp.02}), if and only if,
\begin{equation}
V_{G}^{\left( 4\right) }=V\left( v-\frac{c_{1}}{c_{2}}u,\bar{s},\bar{t}%
\right) .
\end{equation}

\subsection{Isometry class 5}

In this case the spacetime (\ref{pp.01}) admits a three dimensional Killing
algebra with commutators%
\begin{equation}
\left[ k,X_{2}^{\left( 5\right) }\right] =0~,~\left[ X_{2}^{\left( 5\right)
},X_{3}^{\left( 5\right) }\right] =0~,~\left[ k,X_{3}^{\left( 5\right) }%
\right] =-k,
\end{equation}%
where $X_{2}^{\left( 5\right) }=\partial _{\theta }$, $X_{3}^{\left(
5\right) }=u\partial _{u}-v\partial _{v}$ and $H=u^{-2}W\left( r\right) $.

Therefore, the generic vector field $X_{5}=c_{1}k+c_{2}X_{2}^{\left(
5\right) }+c_{3}X_{3}^{\left( 5\right) }$ is a Lie point symmetry of (\ref%
{pp.02}), if and only if%
\begin{equation}
V_{G}^{\left( 5\right) }=V\left( v-\frac{c_{1}}{c_{3}}u,~r~,~\theta -\frac{%
c_{2}}{c_{3}}\ln \left( u\right) \right) .
\end{equation}

Moreover, for the subclasses (5i) with $W\left( r\right) =\zeta \ln r,$ and
(5ii) with $W\left( r\right) =\left( \delta r^{-\sigma }-\sigma \left(
2-\sigma \right) ^{-2}r^{2}\right) ,~\left\vert \sigma \right\vert \neq 0,2$%
, the spacetime (\ref{pp.01}) admits a four dimensional conformal algebra.

\subsubsection{Subclass 5i}

When $H=\zeta u^{-2}\ln r$ the spacetime (\ref{pp.01}) admits the extra
sp.CKV
\begin{equation}
S_{4}^{\left( 5i\right) }=u^{2}\partial _{u}+\left( \frac{r^{2}}{2}-\zeta
\ln u\right) \partial _{v}+ur\partial _{r}~,~\psi _{4}^{\left( 5i\right) }=u.
\end{equation}%
Hence the field $Y_{4}^{\left( 5i\right) }=S_{4}^{\left( 5i\right) }-\frac{1%
}{2}uX_{\Psi }$ is a Lie point symmetry of (\ref{pp.02}) provided%
\begin{equation}
V_{4}^{\left( 5i\right) }=u^{-2}V\left( ru^{-1},v-\frac{r^{2}+2\zeta \left(
1+\ln u\right) }{2u},\theta \right) .
\end{equation}

Similarly the field $X_{\left( 5i\right) }=c_{1}k+c_{2}X_{2}^{\left(
5\right) }+c_{3}X_{3}^{\left( 5\right) }+c_{4}Y_{4}^{\left( 5i\right) }$ is
a Lie point symmetry of (\ref{pp.02}) when%
\begin{equation}
V_{G}^{\left( 5i\right) }=\left( c_{3}+c_{4}u\right) ^{-2}V\left( \frac{r}{%
c_{3}+c_{4}u},F\left( u,r,v\right) ,\theta -\frac{c_{2}}{c_{3}}\ln \left(
\frac{u}{c_{3}+c_{4}u}\right) \right) ,
\end{equation}%
where
\begin{equation}
F\left( u,r,v\right) =\frac{c_{1}+c_{4}uv}{c_{4}\left( c_{3}+c_{4}u\right) }-%
\frac{c_{4}ur^{2}}{\left( c_{3}+c_{4}u\right) ^{2}}-\frac{\zeta }{c_{3}}\ln
\left( c_{3}+c_{4}u\right) +\frac{c_{4}}{c_{3}}\frac{\zeta \left( 1+u\ln
u\right) }{\left( c_{3}+c_{4}u\right) }.
\end{equation}

\subsubsection{Subclass 5ii}

Contrary to the subclass (5i), this subclass admits the proper CKV%
\begin{equation}
C_{4}^{\left( 5ii\right) }=u^{\frac{4}{2-\sigma }}\partial _{u}+\frac{\left(
\sigma +2\right) }{\left( \sigma -2\right) ^{2}}r^{2}u^{\frac{2\sigma }{%
2-\sigma }}\partial _{v}+\frac{2}{2-\sigma }ru^{\frac{\sigma +2}{2-\sigma }%
}\partial _{r},
\end{equation}%
with conformal factor $\psi _{4}^{\left( 5ii\right) }=\frac{2}{\sigma -2}u^{%
\frac{\sigma +2}{2-\sigma }}$. Since $C_{4}^{\left( 5ii\right) }$ is a not a
sp.CKV holds that $\left( \psi _{4}^{\left( 5ii\right) }\right) _{;ij}\neq 0$%
; however we note that $\Delta \left( \psi _{4}^{\left( 5ii\right) }\right)
=0$ which means that $\psi _{4}^{\left( 5ii\right) }$ is a solution of the
wave equation. Therefore, we have that the vector field $Y_{4}^{\left(
5ii\right) }=C_{4}^{\left( 5ii\right) }-\frac{1}{2}\psi _{4}^{\left(
5ii\right) }$ is a point symmetry of (\ref{pp.02}) when
\begin{equation}
V_{4}^{\left( 5ii\right) }=u^{\frac{4}{\sigma -2}}V\left( ru^{\frac{2}{%
\sigma -2}},v+\frac{r^{2}}{u\left( \sigma -2\right) },\theta \right) ,
\end{equation}

Furthermore, the field $Y_{\left( 5ii\right) }=c_{1}k+c_{2}X_{2}^{\left(
5\right) }+c_{3}X_{3}^{\left( 5\right) }+Y_{4}^{\left( 5ii\right) }$ is a
point symmetry of (\ref{pp.02}) when%
\begin{equation}
V_{G}^{\left( 5ii\right) }=\left( f_{1}\left( u\right) \right) ^{2}V\left(
rf_{1}\left( u\right) ,g\left( v,u,r\right) ,\theta -\frac{c_{2}}{c_{3}}\ln
f_{1}\left( u\right) \right) ,
\end{equation}%
where%
\begin{equation}
g\left( v,u,r\right) =v\left( f_{2}\left( u\right) \right) ^{\frac{\sigma -2%
}{\sigma +2}}-\frac{c_{1}}{c_{3}}\left( f_{2}\left( u\right) \right) ^{-%
\frac{4}{\sigma +2}}f_{2}\left( u\right) +c_{4}\frac{u^{\frac{4}{\sigma +2}}%
}{2-\sigma }\left( f_{1}\left( u\right) \right) ^{2}f_{3}\left( u\right)
r^{2},
\end{equation}%
and%
\begin{equation}
f_{1}\left( u\right) =\left( c_{3}+c_{4}u^{\frac{\sigma +2}{2-\sigma }%
}\right) ^{-\frac{2}{\sigma +2}}~,~f_{2}\left( u\right) =\left(
c_{4}+c_{3}u^{\frac{\sigma +2}{2-\sigma }}\right) ,
\end{equation}%
\begin{equation}
f_{3}\left( u\right) =\int \left( f_{1}\left( u\right) \right) ^{2}\left(
f_{2}\left( u\right) \right) ^{-\frac{4}{2+\sigma }}u^{-\frac{2\left( \sigma
+4\right) }{2+\sigma }}du.
\end{equation}

The commutators of the elements of the conformal algebras of the isometry
class 5 with the subclasses (5i) and (5ii) are given in table \ref{comClass5}%
.

\begin{table}[tbp] \centering%
\caption{Commutators of the conformal algebras of isometry class 5 of
pp-wave spacetime (\ref{pp.01}) and of the subclasses (5i), (5ii)}%
\begin{tabular}{c|ccccc}
\hline\hline
$\left[ \mathbf{X}_{I},\mathbf{X}_{J}\right] $ & $k$ & $X_{2}^{\left(
5\right) }$ & $X_{3}^{\left( 5\right) }$ & $S_{4}^{\left( 5i\right) }$ & $%
C_{4}^{\left( 5ii\right) }$ \\ \hline
$k$ & $0$ & $0$ & $-k$ & $0$ & $0$ \\
$X_{2}^{\left( 5\right) }$ & $0$ & $0$ & $0$ & $0$ & $0$ \\
$X_{3}^{\left( 5\right) }$ & $k$ & $0$ & $0$ & $S_{4}^{\left( 5i\right)
}-\zeta k$ & $-\frac{4}{\left( \sigma -2\right) }C_{4}^{\left( 5ii\right) }$
\\
$S_{4}^{\left( 5i\right) }$ & $0$ & $0$ & $-S_{4}^{\left( 5i\right) }+\zeta
k $ & $0$ &  \\
$C_{4}^{\left( 5ii\right) }$ & $0$ & $0$ & $\frac{4}{\left( \sigma -2\right)
}C_{4}^{\left( 5ii\right) }$ &  & $0$ \\ \hline\hline
\end{tabular}%
\label{comClass5}%
\end{table}%

\subsection{Isometry class 6}

In the isometry class 6, the spacetime (\ref{pp.01}) admits three KVs, the
field $k$ and the two vector fields $X_{2}^{\left( 6\right) }=\partial
_{\theta }~,~X_{3}^{\left( 6\right) }=\partial _{u},$ where $H\left(
u,x^{A}\right) =W\left( r\right) $.

Hence, we have that the generic field $X_{6}=c_{1}k+c_{2}X_{2}^{\left(
6\right) }+c_{3}X_{3}^{\left( 6\right) },$ is a Lie point symmetry of (\ref%
{pp.02}), if%
\begin{equation}
V_{G}^{\left( 6\right) }=V\left( v-\frac{c_{1}}{c_{3}}u,r,\theta -\frac{c_{2}%
}{c_{3}}u\right) .
\end{equation}

For special functions $H\left( r\right) $ the spacetime (\ref{pp.01}) admits
a greater conformal algebra. There exist four possible subclasses in which
the pp-wave spacetime admits a greater conformal algebra.

\subsubsection{Subclass 6i}

When $W\left( r\right) =\frac{N}{4}r^{2}+\delta r^{-2}$, the spacetime (\ref%
{pp.01}) admits two proper non sp.CKVs, the fields%
\begin{eqnarray}
C_{4}^{\left( 6i\right) } &=&\sin \left( \sqrt{2N}u\right) \left( \partial
_{u}-\frac{Nr^{2}}{2}\partial _{v}\right) +\frac{\sqrt{2N}}{2}r\cos \left(
\sqrt{2N}u\right) \partial _{r}, \\
C_{5}^{\left( 6i\right) } &=&\cos \left( \sqrt{2N}u\right) \left( \partial
_{u}-\frac{Nr^{2}}{2}\partial _{v}\right) -\frac{\sqrt{2N}}{2}r\sin \left(
\sqrt{2N}u\right) \partial _{r},
\end{eqnarray}%
with conformal factors $\psi _{4}^{\left( 6i\right) }=\frac{\sqrt{2N}}{2}%
\cos \left( \sqrt{2N}u\right) $ and $\psi _{5}^{\left( 6i\right) }=-\frac{%
\sqrt{2N}}{2}\sin \left( \sqrt{2N}u\right) $ which satisfy the wave
equation, i.e. $\Delta \psi _{4-5}^{\left( 6i\right) }=0$.

Therefore, from the fields $C_{4}^{\left( 6i\right) },$ $C_{5}^{\left(
6i\right) }$ we have the possible point symmetries $Y_{4}^{\left( 6i\right)
}=C_{4}^{\left( 6i\right) }-\frac{1}{2}\psi _{4}^{\left( 6i\right) }X_{\Psi
} $ and $Y_{5}^{\left( 6i\right) }=C_{5}^{\left( 6i\right) }-\frac{1}{2}\psi
_{5}^{\left( 6i\right) }X_{\Psi }$ for the Klein-Gordon equation (\ref{pp.02}%
) provided the potential has the following forms%
\[
Y_{4}^{\left( 6i\right) }:V_{4}^{\left( 6i\right) }=\sin \left( \sqrt{2N}%
u\right) V\left( r^{2}\sin \left( \sqrt{2N}u\right) ,v-\frac{\sqrt{2N}}{4}%
r^{2}\cot \left( \sqrt{2N}u\right) ,\theta \right) ,
\]%
\[
Y_{5}^{\left( 6i\right) }:V_{5}^{\left( 6i\right) }=\cos \left( \sqrt{2N}%
u\right) V\left( r^{2}\cos \left( \sqrt{2N}u\right) ,v+\frac{\sqrt{2N}}{4}%
r^{2}\tan \left( \sqrt{2N}u\right) ,\theta \right) .
\]

Moreover, if the general vector field $X_{\left( 6i\right)
}=X_{6}+c_{4}Y_{4}^{\left( 6i\right) }+c_{5}Y_{5}^{\left( 6i\right) }$ is a
point symmetry of (\ref{pp.02}), then%
\[
V_{G}^{\left( 6i\right) }=V\left( g_{1},g_{2},g_{3}\right) ,
\]%
where%
\begin{equation}
g_{1}=r^{2}f_{1}\left( u\right) ~,~g_{3}=\theta -c_{2}f_{2}\left( u\right) ,
\end{equation}%
\[
g_{2}=2v+\sqrt{2N}g_{1}\left( \frac{1}{2}c_{5}\sin \left( \sqrt{2N}u\right)
-c_{4}\cos ^{2}\left( \frac{\sqrt{2N}}{2}u\right) \right) -2c_{1}f_{2}\left(
u\right) ,
\]%
and%
\begin{eqnarray}
f_{1}\left( u\right) &=&\left( c_{3}+c_{4}\sin \left( \sqrt{2N}u\right)
+c_{5}\cos \left( \sqrt{2N}u\right) \right) ^{-1} \\
f_{2}\left( u\right) &=&\frac{\sqrt{2}}{\sqrt{N}\sqrt{\left( c_{3}\right)
^{2}-\left( c_{4}\right) ^{2}-\left( c_{5}\right) ^{2}}}\arctan \left( \frac{%
\left( c_{3}-c_{5}\right) \tan \left( \frac{\sqrt{2N}}{2}u\right) +c_{4}}{%
\sqrt{\left( c_{3}\right) ^{2}-\left( c_{4}\right) ^{2}-\left( c_{5}\right)
^{2}}}\right) .
\end{eqnarray}

\subsubsection{Subclass 6ii}

When $N=0$, i.e. $W\left( r\right) =\delta r^{-2}$, the spacetime (\ref%
{pp.01}) admits a HV and a proper sp.CKV. These fields are respectively
\begin{equation}
H_{4}^{\left( 6ii\right) }=2u\partial _{u}+r\partial _{r}~,~\psi
_{4}^{\left( 6ii\right) }=1,
\end{equation}%
\begin{equation}
S_{5}^{\left( 6ii\right) }=u^{2}\partial _{u}+\frac{r^{2}}{2}\partial
_{v}+ru\partial _{r}~,~\psi _{5}^{\left( 6ii\right) }=\frac{u}{2}.
\end{equation}%
Note that $\left( \psi _{5}^{\left( 6ii\right) }\right) _{;\mu \nu }=0$.
Therefore from the fields $H_{4}^{\left( 6ii\right) }$ and $S_{5}^{\left(
6ii\right) }$ we have the possible Lie point symmetries of (\ref{pp.02}) $%
Y_{4}^{\left( 6ii\right) }=H_{4}^{\left( 6ii\right) }-\frac{1}{2}X_{\Psi }$
and $Y_{5}^{\left( 6ii\right) }=S_{5}^{\left( 6ii\right) }-\frac{1}{2}\psi
_{5}^{\left( 6ii\right) }X_{\Psi }$ respectively.

This means that the generic point symmetry vector of (\ref{pp.02}) is $%
X_{\left( 6ii\right) }=X_{\left( 6\right) }+c_{4}Y_{4}^{\left( 6ii\right)
}+c_{5}Y_{5}^{\left( 6ii\right) }$ provided the potential has the form%
\begin{equation}
V_{G}^{\left( 6ii\right) }=f_{1}\left( u\right) V\left( r^{2}f_{1}\left(
u\right) ,v-\frac{c_{5}}{2}r^{2}f_{1}\left( u\right) -c_{1}f_{2}\left(
u\right) ,\theta -c_{2}f_{2}\right) ,
\end{equation}%
where
\begin{equation}
f_{1}\left( u\right) =\left( c_{3}+2c_{4}u+c_{5}u^{2}\right)
^{-1}~~,~f_{2}\left( u\right) =\frac{\arctan \left( \frac{c_{4}+c_{5}u}{%
\sqrt{c_{3}c_{5}-\left( c_{4}\right) ^{2}}}\right) }{\sqrt{c_{3}c_{5}-\left(
c_{4}\right) ^{2}}}.
\end{equation}

\subsubsection{Subclass 6iii}

When $W\left( r\right) =\zeta \ln r$, the spacetime (\ref{pp.01}) admits the
extra HV%
\begin{equation}
H_{4}^{\left( 6iii\right) }=u\partial _{u}+\left( v-\zeta u\right) \partial
_{v}+r\partial _{r}~,~\psi _{4}^{\left( 6i\right) }=1.
\end{equation}%
It follows that $Y_{4}^{\left( 6iii\right) }=H_{4}^{\left( 6iii\right) }-%
\frac{1}{2}X_{\Psi }$ is a Lie point symmetry of (\ref{pp.02}) when
\begin{equation}
V=\frac{1}{u^{2}}V\left( vu^{-1}+\zeta \ln u,ru^{-1},\theta \right) .
\end{equation}

Moreover, the vector field $X_{\left( 6iii\right) }=X_{6}+c_{4}Y_{4}^{\left(
6iii\right) }$ is a Lie and Noether point symmetry of (\ref{pp.02}) if and
only if the potential has the following form
\begin{equation}
V_{G}^{\left( 6iii\right) }=V\left( \frac{c_{4}v+c_{1}+c_{3}\zeta }{%
c_{4}\left( c_{3}+c_{4}u\right) }-\frac{\zeta }{c_{4}}\ln \left(
c_{3}+c_{4}u\right) ,\frac{r}{c_{3}+c_{4}u},\theta -\frac{c_{2}}{c_{4}}\ln
\left( c_{3}+c_{4}u\right) \right) .
\end{equation}

\subsubsection{Subclass 6iv}

When $W\left( r\right) =\delta r^{-s}$ with $s\neq 2,0$, $\ $the spacetime (%
\ref{pp.01}) admits the extra HV%
\begin{equation}
H_{4}^{\left( 6iv\right) }=\frac{2+\sigma }{2}u\partial _{u}+\frac{2-\sigma
}{2}v\partial _{v}+r\partial _{r}~,~\psi _{4}^{\left( 6iv\right) }.
\end{equation}

Hence, the vector field $X_{\left( 6iv\right) }=X_{6}+c_{4}\left(
H_{5}^{\left( 6iv\right) }-\frac{1}{2}X_{\Psi }\right) $ is a Lie and
Noether point symmetry of (\ref{pp.02}) when
\begin{equation}
V_{G}^{\left( 6iv\right) }=\left( f_{1}\left( u\right) \right) \left( \left(
v+2\frac{c_{1}}{c_{3}\left( 2-\sigma \right) }\right) ^{2}f_{1}\left(
u\right) ,r^{2}f_{1}\left( u\right) ,2\theta +c_{2}\ln f_{1}\left( u\right)
\right) .
\end{equation}%
where $f_{1}\left( u\right) =\left( 2c_{3}+c_{4}u\left( 2+\sigma \right)
u\right) ^{-\frac{4}{2+\sigma }}$.

In table \ref{comClass6}, we give the commutators of the vector fields which
form the conformal algebras of the isometry class 6 and the subclasses
(6i)-(6iv).

\begin{table}[tbp] \centering%
\caption{Commutators of the conformal algebras of isometry class 6 of
spacetime (\ref{pp.01}) and of the subclasses (6i)-(6iv)}%
\begin{tabular}{c|ccccccccc}
\hline\hline
$\left[ \mathbf{X}_{I},\mathbf{X}_{J}\right] $ & $k$ & $X_{2}^{\left(
6\right) }$ & $X_{3}^{\left( 6\right) }$ & $C_{4}^{\left( 6i\right) }$ & $%
C_{5}^{\left( 6i\right) }$ & $H_{4}^{\left( 6ii\right) }$ & $S_{5}^{\left(
6ii\right) }$ & $H_{4}^{\left( 6iii\right) }$ & $H_{4}^{\left( 6iv\right) }$
\\ \hline
$k$ & $0$ & $0$ & $0$ & $0$ & $0$ & $0$ & $0$ & $k$ & $\left( 1-\frac{\sigma
}{2}\right) k$ \\
$X_{2}^{\left( 6\right) }$ & $0$ & $0$ & $0$ & $0$ & $0$ & $0$ & $0$ & $0$ &
$0$ \\
$X_{3}^{\left( 6\right) }$ & $0$ & $0$ & $0$ & $C_{5}^{\left( 6i\right) }$ &
$-C_{4}^{\left( 6i\right) }$ & $2X_{3}^{\left( 6\right) }$ & $H_{4}^{\left(
6ii\right) }$ & $X_{3}^{\left( 6\right) }-\zeta k$ & $\left( 1+\frac{\sigma
}{2}\right) X_{3}^{\left( 6\right) }$ \\
$C_{4}^{\left( 6i\right) }$ & $0$ & $0$ & $-C_{5}^{\left( 6i\right) }$ & $0$
& $-X_{3}^{\left( 6\right) }$ &  &  &  &  \\
$C_{5}^{\left( 6i\right) }$ & $0$ & $0$ & $C_{4}^{\left( 6i\right) }$ & $%
X_{3}^{\left( 6\right) }$ & $0$ &  &  &  &  \\
$H_{4}^{\left( 6ii\right) }$ & $0$ & $0$ & $-2X_{3}^{\left( 6\right) }$ &  &
& $0$ & $2S_{6}^{\left( 6ii\right) }$ &  &  \\
$S_{5}^{\left( 6ii\right) }$ & $0$ & $0$ & $-H_{4}^{\left( 6ii\right) }$ &
&  & $2S_{6}^{\left( 6ii\right) }$ & $0$ &  &  \\
$H_{4}^{\left( 6iii\right) }$ & $-k$ & $0$ & $-X_{3}^{\left( 6\right)
}+\zeta k$ &  &  &  &  & $0$ &  \\
$H_{4}^{\left( 6iv\right) }$ & $\left( \frac{\sigma }{2}-1\right) k$ & $0$ &
$-\left( 1+\frac{\sigma }{2}\right) X_{3}^{\left( 6\right) }$ &  &  &  &  &
& $0$ \\ \hline\hline
\end{tabular}%
\label{comClass6}%
\end{table}%

\subsection{Isometry class 7}

In the isometry class 7 $H\left( u,x^{A}\right) =e^{2\omega \theta }W\left(
r\right) .$ For this $H\left( u,x^{A}\right) $ spacetime (\ref{pp.01})
admits as KVs the fields $k$ and
\begin{equation}
X_{2}^{\left( 7\right) }=\omega u\partial _{u}-\omega v\partial
_{v}-\partial _{\theta }~,~X_{3}^{\left( 7\right) }=\partial _{u}.
\end{equation}%
The commutators of the Killing algebra are%
\[
\left[ k,X_{2}^{\left( 7\right) }\right] =-\omega k~,~\left[ k,X_{3}^{\left(
7\right) }\right] =0~,~\left[ X_{2}^{\left( 7\right) },X_{3}^{\left(
7\right) }\right] =-\omega X_{3}^{\left( 7\right) }.
\]

The field $X_{2}^{\left( 7\right) }$ is a point symmetry of (\ref{pp.02})
when
\begin{equation}
V_{2}^{\left( 7\right) }=V\left( vu,r,\omega \theta +\ln u\right) .
\end{equation}

Moreover, the generic KV $\ X_{7}=c_{1}k+c_{2}X_{2}^{\left( 7\right)
}+c_{3}X_{3}^{\left( 7\right) }$, is a \ Lie symmetry for equation (\ref%
{pp.02}) if and only if,%
\begin{equation}
V_{G}^{\left( 7\right) }=V\left( v\left( c_{3}+c_{2}\omega u\right)
-c_{1}u,r,\omega \theta +\ln \left( c_{2}\omega u+c_{3}\right) \right) .
\end{equation}

We note, that there exist subclasses of the isometry class 7 in which the
spacetime (\ref{pp.01}) admits a greater conformal algebra; however, these
subclasses are not vacuum pp-wave spacetimes. We continue with the isometry
class 8.

\subsection{Isometry class 8}

Consider $H=\exp \left( 2\delta t^{\prime }\right) W\left( s^{\prime
}\right) $ where
\begin{equation}
t^{\prime }=\eta y+\sigma z~,~s^{\prime }=\eta z-\sigma y,
\end{equation}%
$\delta =-\frac{c}{\eta ^{2}+\sigma ^{2}}$ and $c,\eta ,\sigma \in
\mathbb{R}
-\left\{ \zeta ^{2}\equiv \eta ^{2}+\sigma ^{2}=0\right\} $.

For $\delta \neq 0$, the spacetime (\ref{pp.01}) admits the three KVs%
\begin{equation}
k,~X_{2}^{\left( 8\right) }=u\partial _{u}~,~X_{3}^{\left( 8\right) }=\delta
\left( u\partial _{u}-v\partial _{v}\right) -\partial _{t^{\prime }},
\end{equation}%
and for $\delta =0$ admits four KVs, the extra KV is%
\[
X_{4}^{\left( 8\right) }=t^{\prime }\partial _{v}+\left( \eta ^{2}+\sigma
^{2}\right) u\partial _{t^{\prime }}.
\]

Therefore, for $\delta \neq 0$, the general form of the potential $V\left(
u,v,s^{\prime },t^{\prime }\right) $ for which the Klein-Gordon equation
admits as a Lie point symmetry the vector field~$X_{8}=c_{1}k+c_{2}X_{2}^{%
\left( 8\right) }+c_{3}X_{3}^{\left( 8\right) },$ is as follows:%
\begin{equation}
V_{G}^{\left( 8\right) }=V\left( c_{1}u+v\left( \delta c_{3}u-c_{2}\right)
,s^{\prime },\delta t^{\prime }+\ln \left( \delta c_{3}u-c_{2}\right)
\right) .
\end{equation}

Finally, the commutators of the Killing algebra are as follows%
\[
\left[ k,X_{2}^{\left( 8\right) }\right] =0~,~\left[ k,X_{3}^{\left(
8\right) }\right] =\delta k,~\left[ X_{2}^{\left( 8\right) },X_{3}^{\left(
8\right) }\right] =-X_{2}^{\left( 8\right) }.
\]

Similarly, for $\delta =0$ we have that the vector field $X_{8}^{\left(
0\right) }=X_{8}+c_{3}X_{4}^{\left( 8\right) }$ is a Lie point symmetry of (%
\ref{pp.02}), if and only if
\begin{equation}
V_{G\left( 0\right) }^{\left( 8\right) }=V\left(
\begin{array}{c}
s^{\prime },c_{2}t^{\prime }-\left[ c_{3}+\frac{c_{4}\zeta ^{2}}{2}u\right]
u, \\
c_{4}t^{\prime }u-c_{2}v+u\left( c_{1}u-\frac{c_{3}c_{4}}{2c_{2}}u^{2}-\frac{%
c_{4}^{2}\zeta ^{2}}{3c_{2}}u^{2}\right)%
\end{array}%
\right) .
\end{equation}

\subsubsection{Subclass 8i}

When $\delta =0$ and $W\left( s\right) =Ks^{\gamma }$, $K,\gamma \in
\mathbb{R}
^{\ast }$ and $\gamma \neq \pm 2,1$, the spacetime\ (\ref{pp.01}) admits a
five dimensional homothetic algebra. The extra HV is%
\begin{equation}
H_{5}^{\left( 8i\right) }=\left( 1-\frac{\gamma }{2}\right) u\partial
_{u}+\left( 1+\frac{\gamma }{2}\right) v\partial _{v}+s^{\prime }\partial
_{s^{\prime }}+t^{\prime }\partial _{t^{\prime },}
\end{equation}%
where $\psi _{5}^{\left( 8i\right) }=1$. Hence, the field $Y_{5}^{\left(
8i\right) }=H_{5}^{\left( 8i\right) }-\frac{1}{2}X_{\Psi }$ is a Lie point
symmetry of (\ref{pp.02}) when%
\begin{equation}
V_{5}^{\left( 8i\right) }=u^{\frac{4}{\gamma -2}}V\left( vu^{\frac{\gamma +2%
}{\gamma -2}},s^{\prime }u^{\frac{2}{\gamma -2}},t^{\prime }u^{\frac{2}{%
\gamma -2}}\right) ,~\gamma \neq 2.
\end{equation}

Finally the generic field $X_{\left( 8i\right) }=X_{8}^{\left( 0\right)
}+c_{5}Y_{5}^{\left( 8i\right) }$, is a Lie point symmetry for equation (\ref%
{pp.02})), when%
\begin{equation}
V_{G}^{\left( 8i\right) }=V\left( s^{\prime }\left( g_{1}\right) ^{\frac{2}{%
\gamma -2}},\left( g_{1}\right) ^{\frac{2}{\gamma -2}}\left[ \gamma
c_{5}\left( c_{5}t^{\prime }+c_{3}\right) +2c_{4}\left( \eta ^{2}+\gamma
^{2}\right) \left( c_{2}+c_{5}u\right) \right] ,\frac{\left( g_{1}\right) ^{%
\frac{\gamma +2}{\gamma -2}}g_{2}}{\gamma ^{2}\left( \gamma +2\right) \left(
c_{5}\right) ^{3}}\right) ,
\end{equation}%
where functions $g_{1},g_{2}$ are%
\begin{equation}
g_{1}=c_{5}\left( \gamma -1\right) u-2c_{2},
\end{equation}%
\begin{eqnarray}
g_{2} &=&2c_{1}c_{5}^{2}\gamma ^{2}+4c_{2}\left( c_{4}\right) ^{4}\zeta
^{2}+4\gamma c_{3}c_{4}c_{5}+  \nonumber \\
&&+2\left( \gamma +2\right) c_{5}\left[ c_{4}\left( c_{5}t^{\prime }\gamma
+c_{4}\zeta ^{2}u\right) +\left( c_{5}\right) ^{2}\gamma ^{2}v\right] .
\end{eqnarray}

We continue with the subclasses (8ii) and (8iii), where $\gamma =2$ and $%
\gamma =-2$ respectively. As we have discussed, when $\gamma =1$ the space (%
\ref{pp.01}) is flat, and we do not consider that case.

\subsubsection{Subclass 8ii}

When $\gamma =2$, i.e. $W\left( s\right) =Ks^{2}$, the pp-wave spacetime (%
\ref{pp.01}) admits a seven dimensional conformal algebra. In particular
admits six KVs and a HV. The KVs are the fields $k,~X_{2}^{\left( 8\right)
},~X_{3}^{\left( 8\right) },X_{4}^{\left( 8\right) }~$and
\begin{eqnarray}
X_{5}^{\left( 8ii\right) } &=&\frac{\sqrt{2K}}{\zeta }s^{\prime }\cos \left(
\sqrt{2K}\zeta u\right) \partial _{v}+\sin \left( \sqrt{2K}\zeta u\right)
\partial _{s^{\prime }}, \\
X_{6}^{\left( 8ii\right) } &=&\frac{\sqrt{2K}}{\zeta }s^{\prime }\sin \left(
\sqrt{2K}\zeta u\right) \partial _{v}-\cos \left( \sqrt{2K}\zeta u\right)
\partial _{s^{\prime }},
\end{eqnarray}%
where the HV is the $H_{5}^{\left( 8i\right) }$. \ Therefore, the fields $%
X_{5}^{\left( 8ii\right) }$ and $X_{6}^{\left( 8ii\right) }$ are Lie point
symmetries of (\ref{pp.02}) when
\begin{eqnarray}
X_{5}^{\left( 8ii\right) } &:&V_{5}^{\left( 8ii\right) }=V\left( u,s^{\prime
2}-\frac{\sqrt{2}\zeta }{\sqrt{K}}v\tan \left( \sqrt{2K}\zeta u\right)
,t^{\prime }\right) , \\
X_{6}^{\left( 8ii\right) } &:&V_{6}^{\left( 8ii\right) }=V\left( u,s^{\prime
2}+\frac{\sqrt{2}\zeta }{\sqrt{K}}v\cot \left( \sqrt{2K}\zeta u\right)
,t^{\prime }\right) .
\end{eqnarray}

We continue with the next subclass, where $\gamma =-2.$

\subsubsection{Subclass 8iii}

When $W\left( s\right) =Ks^{-2}$, the pp-wave spacetime (\ref{pp.01}) admits
four KVs, one HV and the sp.CKV%
\begin{equation}
S_{6}^{\left( 8iii\right) }=u^{2}\partial _{u}+\frac{s^{\prime 2}+t^{\prime
2}}{2\zeta ^{2}}\partial _{v}+us^{\prime }\partial _{s^{\prime }}+ut^{\prime
}\partial _{t^{\prime }},~\psi _{6}^{\left( 8iii\right) }=u.
\end{equation}%
Therefore, from the sp.CKV we have that vector field $Y_{6}^{\left(
8iii\right) }=S_{6}^{\left( 8iii\right) }-\frac{1}{2}\psi _{6}^{\left(
8iii\right) }X_{\Psi }$ is a Lie point symmetry of the Klein-Gordon equation
when%
\begin{equation}
V_{6}^{\left( 8iii\right) }=u^{2}V\left( v-\frac{s^{\prime 2}+t^{\prime 2}}{%
2\zeta ^{2}u},su^{-1},tu^{-1}\right) .
\end{equation}

The commutators of the elements of the conformal algebras of the isometry
class 8 for $\delta =0,~$and of the subclasses (8i)-(8iii) are given in
table \ref{comClass8}. We would like to remark that for the subclasses (8ii)
and (8iii), we did not give the form of the potential for which the
Klein-Gordon equation (\ref{pp.02}) admits as Lie point symmetry the generic
symmetry vector, which follows from the linear combination of the CKVs,
because in this case the potential has a complex functional form.

\begin{table}[tbp] \centering%
\caption{Commutators of the conformal algebras for the isometry class 8 for $\delta=0$ and of the subclasses (8i)-(8iii) for
the pp-wave spacetime (\ref{pp.01}).}%
\begin{tabular}{c|cccccccc}
\hline\hline
$\left[ \mathbf{X}_{I},\mathbf{X}_{J}\right] $ & $k$ & $X_{2}^{\left(
8\right) }$ & $X_{3}^{\left( 8\right) }$ & $X_{4}^{\left( 8\right) }$ & $%
H_{5}^{\left( 8i\right) }$ & $X_{5}^{\left( 8ii\right) }$ & $X_{6}^{\left(
8ii\right) }$ & $S_{6}^{\left( 8iii\right) }$ \\ \hline
$k$ & $0$ & $0$ & $0$ & $0$ & $\left( 1+\frac{\gamma }{2}\right) k$ & $0$ & $%
0$ & $0$ \\
$X_{2}^{\left( 8\right) }$ & $0$ & $0$ & $0$ & $\zeta ^{2}X_{3}^{\left(
8\right) }$ & $\left( 1-\frac{\gamma }{2}\right) X_{2}^{\left( 8\right) }$ &
$-\sqrt{2K}\zeta X_{6}^{\left( 8ii\right) }$ & $\sqrt{2K}\zeta X_{5}^{\left(
8ii\right) }$ & $H_{5}^{\left( 8i\right) }$ \\
$X_{3}^{\left( 8\right) }$ & $0$ & $0$ & $0$ & $k$ & $X_{3}^{\left( 8\right)
}$ & $0$ & $0$ & $\frac{X_{4}^{\left( 8\right) }}{\zeta ^{2}}$ \\
$X_{4}^{\left( 8\right) }$ & $0$ & $-\zeta ^{2}X_{3}^{\left( 8\right) }$ & $%
-k$ & $0$ & $\frac{\gamma X_{4}^{\left( 8\right) }}{2}$ & $0$ & $0$ & $0$ \\
$H_{5}^{\left( 8i\right) }$ & $-\left( 1+\frac{\gamma }{2}\right) k$ & $%
\left( \frac{\gamma }{2}-1\right) X_{2}^{\left( 8\right) }$ & $%
-X_{3}^{\left( 8\right) }$ & $\frac{-\gamma X_{4}^{8}}{2}$ & $0$ & $%
-X_{5}^{\left( 8ii\right) }$ & $-X_{6}^{\left( 8ii\right) }$ & $%
2S_{6}^{\left( 8iii\right) }$ \\
$X_{5}^{\left( 8ii\right) }$ & $0$ & $\sqrt{2K}\zeta X_{6}^{\left(
8ii\right) }$ & $0$ & $0$ & $X_{5}^{\left( 8ii\right) }$ & $0$ & $\frac{%
\sqrt{2K}}{\zeta }k$ &  \\
$X_{6}^{\left( 8ii\right) }$ & $0$ & $-\sqrt{2K}\zeta X_{5}^{\left(
8ii\right) }$ & $0$ & $0$ & $X_{6}^{\left( 8ii\right) }$ & $-\frac{\sqrt{2K}%
}{\zeta }k$ & $0$ &  \\
$S_{6}^{\left( 8iii\right) }$ & $0$ & $H_{5}^{\left( 8i\right) }$ & $-\frac{%
X_{4}^{\left( 8\right) }}{\zeta ^{2}}$ & $0$ & $-2S_{6}^{\left( 8iii\right)
} $ &  &  & $0$ \\ \hline\hline
\end{tabular}%
\label{comClass8}%
\end{table}%

\subsection{Isometry class 9}

In isometry class 9, $H=Ke^{\eta y-\sigma z}$ with $K,\eta ,\sigma ~\in
\mathbb{R}
-\left\{ \zeta ^{2}\equiv \eta ^{2}+\sigma ^{2}=0\right\} .$ In this class,
spacetime (\ref{pp.01}) admits a five dimensional Killing algebra. The\ KVs
are the field $k$ and
\[
X_{2}^{\left( 9\right) }=u\partial _{u},~X_{3}^{\left( 9\right) }=u\partial
_{u}-v\partial _{v}+\frac{1}{\sigma }\partial _{z}~,
\]%
\[
X_{4}^{\left( 9\right) }=\left( y+\frac{\eta }{\sigma }z\right) \partial
_{v}+u\partial _{y}+\frac{\eta }{\sigma }u\partial _{z}~,~X_{5}^{\left(
9\right) }=\partial _{y}+\frac{\eta }{\sigma }\partial _{z}.
\]%
The commutators of the Lie algebra are given in table \ref{comClass9}. In
table \ref{pot9} we give the form of the potential $V\left( u,v,x,y\right) $
for which any of the elements of the Killing algebra of (\ref{pp.01}) is a
point symmetry of (\ref{pp.02}).

\begin{table}[tbp] \centering%
\caption{Commutators of the Killing algebra for the isometry class 9 of
pp-wave spacetime (\ref{pp.01}).}%
\begin{tabular}{c|ccccc}
\hline\hline
$\left[ \mathbf{X}_{I},\mathbf{X}_{J}\right] $ & $k$ & $X_{2}^{\left(
9\right) }$ & $X_{3}^{\left( 9\right) }$ & $X_{4}^{\left( 9\right) }$ & $%
X_{5}^{\left( 9\right) }$ \\ \hline
$k$ & $0$ & $0$ & $-k$ & $0$ & $0$ \\
$X_{2}^{\left( 9\right) }$ & $0$ & $0$ & $X_{2}^{\left( 9\right) }$ & $%
X_{5}^{\left( 9\right) }$ & $0$ \\
$X_{3}^{\left( 9\right) }$ & $k$ & $-X_{2}^{\left( 9\right) }$ & $0$ & $%
\frac{\eta }{\sigma ^{2}}k+X_{4}$ & $0$ \\
$X_{4}^{\left( 9\right) }$ & $0$ & $-X_{5}^{\left( 9\right) }$ & $-\frac{%
\eta }{\sigma ^{2}}k-X_{4}$ & $0$ & $-\frac{\eta ^{2}+\sigma ^{2}}{\sigma
^{2}}k$ \\
$X_{5}^{\left( 9\right) }$ & $0$ & $0$ & $0$ & $\frac{\eta ^{2}+\sigma ^{2}}{%
\sigma ^{2}}k$ & $0$ \\ \hline\hline
\end{tabular}%
\label{comClass9}%
\end{table}%

\begin{table}[tbp] \centering%
\caption{Lie symmetries and potentials for the Klein-Gordon (\ref{pp.02}) in
isometry class 9.}%
\begin{tabular}{cc}
\hline\hline
\textbf{Lie Sym.} & $\mathbf{V}\left( u,v,y,z\right) $ \\ \hline
$k$ & $V\left( u,y,z\right) $ \\
$X_{2}^{\left( 9\right) }$ & $V\left( v,y,z\right) $ \\
$X_{3}^{\left( 9\right) }$ & $V\left( vu,y,ue^{-\sigma z}\right) $ \\
$X_{4}^{\left( 9\right) }$ & $V\left( u,z-\frac{\eta }{\sigma }y,2v\sigma
^{2}-\frac{\zeta ^{2}}{u}y-\frac{2\eta \left( \sigma z-\eta y\right) y}{%
2\sigma ^{2}u}\right) $ \\
$X_{5}^{\left( 9\right) }$ & $V\left( u,v,z-\frac{\eta }{\sigma }y\right) $
\\ \hline\hline
\end{tabular}%
\label{pot9}%
\end{table}%

Furthermore, the generic vector field $X_{9}=c_{1}k+c_{2}X_{2}^{\left(
9\right) }+...+c_{5}X_{5}^{\left( 9\right) }$ is a Lie point symmetry of (%
\ref{pp.02}) if and only if
\begin{equation}
V_{G}^{\left( 9\right) }=V\left( g_{1},g_{2},\bar{g}\left(
c_{3}u+c_{2}\right) ^{C_{3}}\right) ,
\end{equation}%
where $C_{3}=\frac{c_{2}c_{4}}{c_{3}\sigma ^{2}}\left[ \zeta
^{2}c_{2}c_{4}-\eta c_{3}\left( c_{3}+\eta c_{5}\right) -\sigma
^{2}c_{3}c_{5}\right] $,
\begin{equation}
g_{1}\left( u,y\right) =\exp \left( c_{3}\left( c_{3}y-c_{4}u\right) \right)
\left( c_{3}u+c_{2}\right) ^{c_{2}c_{4}-c_{3}c_{5}},
\end{equation}%
\begin{equation}
g_{3}\left( u,z\right) =\exp \left( c_{3}\left( c_{3}\sigma z-c_{4}\eta
u\right) \right) \left( c_{3}u+c_{2}\right) ^{\eta \left(
c_{2}c_{4}-c_{3}c_{5}\right) -\left( c_{3}\right) ^{2}},
\end{equation}%
and%
\begin{eqnarray}
\bar{g}\left( u,v,y,z\right) &=&c_{2}v+\left( c_{3}v-c_{4}y-c_{1}\right) u+%
\frac{c_{4}c_{5}}{c_{3}}\left( \frac{\zeta ^{2}}{\sigma ^{2}}\left( u+\frac{%
c_{2}}{c_{3}}\right) \right) +  \nonumber \\
&&+\frac{c_{2}c_{4}}{\sigma ^{2}\left( c_{3}\right) ^{2}}\left[ \eta
c_{3}-c_{4}\zeta ^{2}\left( u+c_{2}\right) \right] +  \nonumber \\
&&+\frac{\left( c_{4}\right) ^{2}}{2\sigma ^{2}c_{3}}\zeta ^{2}u^{2}+\frac{%
c_{4}\eta \left( 1-\sigma z\right) u}{\sigma ^{2}}.
\end{eqnarray}

\subsection{Plane wave spacetime: Isometry class 10}

When the function $H\left( u,x^{A}\right) $ has the form
\begin{equation}
H\left( u,x^{A}\right) =\frac{1}{2}\left( A\left( u\right) y^{2}+C\left(
u\right) z^{2}\right) +B\left( u\right) yz,  \label{pw.01}
\end{equation}%
the spacetime (\ref{pp.01}) is a plane wave spacetime.

The spacetime (\ref{pp.01}) with (\ref{pw.01}) is vacuum when $A\left(
u\right) +C\left( u\right) =0$. Moreover, admits a six dimensional
homothetic algebra. The four KVs are given by the vector field \cite%
{ppKV,ppCKV,ppnull}%
\begin{equation}
X_{a}^{\left( 10\right) }=\left( y\dot{d}_{a}\left( u\right) +z\dot{e}%
_{a}\left( u\right) \right) \partial _{v}+d_{a}\left( u\right) \partial
_{y}+e_{a}\left( u\right) \partial _{z},
\end{equation}%
where $\dot{d}_{a}=\frac{d}{du}\left( d_{a}\right) ,$ and the functions $%
d_{a}\left( u\right) ,e_{a}\left( u\right) $ satisfy the following system of
equations%
\begin{eqnarray}
\ddot{d}_{a}+Cd_{a}+Be_{a} &=&0, \\
\ddot{e}_{a}+Ae_{a}+Bd_{a} &=&0.
\end{eqnarray}

The fifth KV is the field $k,$ and the proper HV is%
\begin{equation}
H_{6}^{\left( 10\right) }=2v\partial _{v}+y\partial _{y}+z\partial
_{z}~,~\psi _{6}^{\left( 10\right) }=0.
\end{equation}

Hence, we have that the fields $X_{a}^{\left( 10\right) }$ are Lie point
symmetries of (\ref{pp.02}), if and only if%
\begin{equation}
V_{a}^{\left( 10\right) }=V\left( u,v-\frac{\dot{e}_{a}}{d_{a}}yz-\frac{y^{2}%
}{2d_{a}^{2}}\left( \dot{d}_{A}d-\dot{e}_{a}e_{a}\right) ,z-\frac{e_{a}}{%
d_{Aa}}y\right) .
\end{equation}

Moreover, from the HV, we have that the field $Y_{6}^{\left( 10\right)
}=H_{6}^{\left( 10\right) }-\frac{1}{2}X_{\Psi }$ is a Lie point symmetry of
(\ref{pp.02}) provided%
\begin{equation}
V_{6}^{\left( 10\right) }=v^{-1}V\left( u,y^{2}v^{-1},z^{2}v^{-1}\right) .
\end{equation}

Finally, the generic field $X_{10}=c_{1}k+c_{a}X_{a}^{\left( 10\right)
}+c_{6}Y_{6}^{\left( 10\right) }$ is a Lie point symmetry of (\ref{pp.02})
when%
\begin{equation}
V_{G}^{\left( 10\right) }=\left( c_{A}d_{a}+c_{5}y\right) ^{-2}V\left(
u,g\left( v,u,y,z\right) ,\frac{c_{5}z+c_{a}e_{a}}{c_{5}\left(
c_{a}d_{a}+c_{5}y\right) }\right) ,
\end{equation}%
where%
\begin{equation}
g\left( v,u,y,z\right) =\frac{2c_{5}v+c_{1}}{\left( c_{a}d_{a}+c_{5}y\right)
^{2}}+c_{2}\frac{\dot{d}_{a}\left( c_{a}d_{a}+c_{5}y\right) +\dot{e}%
_{A}\left( c_{a}e_{a}+2c_{5}z\right) }{\left( c_{a}d_{a}+c_{5}y\right) ^{2}}.
\end{equation}

The commutators of the homothetic algebra are \cite{ppCKV}%
\begin{eqnarray}
\left[ k,X_{a}^{\left( 10\right) }\right] &=&0~,~\left[ k,H_{6}^{\left(
10\right) }\right] =2k~,~ \\
\left[ X_{a}^{\left( 10\right) },H_{6}^{\left( 10\right) }\right] &=&X_{a}~,~%
\left[ X_{a}^{\left( 10\right) },X_{b}^{\left( 10\right) }\right] =2Q_{\left[
ab\right] }k,
\end{eqnarray}%
where $Q_{ab}\ $are constants. Furthermore, there are three subclasses, for
special form of the functions $A\left( u\right) ,B\left( u\right) ,C\left(
u\right) $ of (\ref{pw.01}), for which the plane symmetric spacetime admits
a greater conformal algebra. In the following we consider the two subclasses
for which the spacetime (\ref{pp.01}) admits extra sp.CKV.

\subsubsection{Subclass 10i}

When the functions $A\left( u\right) ,B\left( u\right) $ and $C\left(
u\right) $ of (\ref{pw.01}) are%
\begin{eqnarray}
A\left( u\right) &=&K\left( u^{2}+\beta \right) ^{-2}\left( \sin \left( \phi
\left( u\right) \right) +\lambda \right) , \\
B\left( u\right) &=&K\left( u^{2}+\beta \right) ^{-2}\cos \left( \phi \left(
u\right) \right) , \\
C\left( u\right) &=&-K\left( u^{2}+\beta \right) ^{-2}\left( \sin \left(
\phi \left( u\right) \right) -\lambda \right) ,
\end{eqnarray}%
where $\phi \left( u\right) =2\gamma \int \frac{du}{u^{2}+\beta },~$the
spacetime (\ref{pp.01}) admits the extra sp.CKV%
\begin{equation}
S_{7}^{\left( 10i\right) }=\left( u^{2}+\beta \right) \partial _{u}+\frac{1}{%
2}\left( y^{2}+z^{2}\right) \partial _{v}+\left( uy+\gamma z\right) \partial
_{y}+\left( uz-\gamma z\right) \partial _{z},
\end{equation}%
where $\psi _{7}^{\left( 10i\right) }=u$. Since $S_{7}^{\left( 10i\right) }$
is a sp.CKV the $\left( \psi _{7}^{\left( 10i\right) }\right) _{;\mu \nu
}=0. $ Therefore we from $S_{7}^{\left( 10i\right) }$ we have that the
vector field $Y_{7}^{\left( 10i\right) }=S_{7}^{\left( 10i\right) }-\frac{1}{%
2}\psi _{7}^{\left( 10i\right) }X_{\Psi }$, is a point symmetry of (\ref%
{pp.02}) when%
\begin{equation}
V_{7}^{\left( 10i\right) }=\left( u^{2}+\beta \right) ^{-1}V\left( v-\frac{%
r^{2}u}{2\left( u^{2}+\beta \right) },\frac{r^{2}}{u^{2}+\beta },\frac{%
e^{2\theta }}{\left( u^{2}+\beta \right) ^{\gamma }}\right) .
\end{equation}

\subsubsection{Subclass 10ii}

When $A\left( u\right) =-\alpha \left( u^{2}+\beta \right) ^{-2}$,~$B\left(
u\right) =-b\left( u^{2}+\beta \right) ^{-2}$ and $C\left( u\right)
=-c\left( u^{2}+\beta \right) ^{-2}$, the extra sp.CKV of (\ref{pp.02}) is%
\begin{equation}
S_{7}^{\left( 10ii\right) }=\left( u^{2}+\beta \right) \partial _{u}+\frac{1%
}{2}\left( y^{2}+z^{2}\right) \partial _{v}+uy\partial _{y}+uz\partial
_{z}~,~\psi _{7}^{\left( 10ii\right) }=u,
\end{equation}%
which is the field $S_{7}^{\left( 10i\right) }$ for $\gamma =0$. Hence, the
field $Y_{7}^{\left( 10ii\right) }=S_{7}^{\left( 10ii\right) }-\frac{1}{2}%
\psi _{7}^{\left( 10ii\right) }X_{\Psi }$ is a Lie point symmetry of
integral when the potential has the form%
\begin{equation}
V_{7}^{\left( 10ii\right) }=\left( u^{2}+\beta \right) ^{-1}V\left( v-\frac{%
r^{2}u}{2\left( u^{2}+\beta \right) },\frac{r^{2}}{u^{2}+\beta },\theta
\right) .
\end{equation}

There are also four more isometry classes in which the plane wave spacetime (%
\ref{pp.01}) admits a seven dimensional homothetic algebra \cite{ppCKV},
where there exists a six dimensional subalgebra and it is the homothetic
algebra of isometry class 10.

For these isometry classes, in table\footnote{%
In the isometry class 12, $\phi =2\delta \ln u.$} \ref{subclasses11}, we
give the functional form of $A,B,C$, the extra KV and the form of the
potential for which the corresponding KV is a Lie point symmetry of the
Klein-Gordon equation (\ref{pp.02}).

\begin{table}[tbp] \centering%
\caption{Point symmetries and potentials for the Klein-Gordon equation
(\ref{pp.02}) in the isometry classes 11-14}%
\begin{tabular}{cccccc}
\hline\hline
\textbf{Class} & $\mathbf{A}\left( u\right) $ & $\mathbf{B}\left( u\right) $
& $\mathbf{C}\left( u\right) $ & \textbf{Extra KV} & \textbf{Potential} \\
\hline
\textbf{11} & $\alpha u^{-2}$ & $\beta u^{-2}$ & $\gamma u^{-2}$ & $%
X_{7}^{\left( 11\right) }=u\partial _{u}-v\partial _{v}$ & $V\left(
vu,y,z\right) $ \\
\textbf{12} & $-cu^{-2}\left( \sin \phi +\lambda \right) $ & $cu^{-2}\left(
\cos \phi \right) $ & $cu^{-2}\left( \sin \phi -\lambda \right) $ & $%
X_{7}^{\left( 12\right) }=X_{7}^{\left( 11\right) }+\delta \partial _{\theta
}$ & $V\left( vu,r,e^{\theta }u^{-\delta }\right) $ \\
\textbf{13} & $\alpha $ & $\beta $ & $c$ & $X_{7}^{\left( 13\right)
}=\partial _{u}$ & $V\left( v,y,z\right) $ \\
\textbf{14} & $-c\sin \left( 2\delta u\right) +\lambda $ & $-c\cos \left(
2\delta u\right) $ & $c\sin \left( 2\delta u\right) +\lambda $ & $%
X_{7}^{\left( 14\right) }=\partial _{u}$ & $V\left( u,r,e^{\theta
}u^{-\delta }\right) $ \\ \hline\hline
\end{tabular}%
\label{subclasses11}%
\end{table}%

\section{Symmetry classification for the Wave equation}

\label{wave}

When the potential in (\ref{pp.02}) vanishes the Klein Gordon equation
becomes
\begin{equation}
-2\Psi _{,uv}+2H\left( u,x^{A}\right) \Psi _{,vv}+\Delta _{\delta }\Psi =0,
\label{wave.01}
\end{equation}%
which is the wave equation in spacetime (\ref{pp.01}).

Contrary to the Klein-Gordon equation, a CKV of the metric which defines the
Laplace operator generates a Lie/Noether symmetry for the wave equation only
when the conformal factor is a solution of the original equation (see
condition (\ref{kg.04})). Therefore, the KVs, the HV and the sp.CKVs
generate always point symmetries for the wave equation. Furthermore, in
section \ref{isometryclass} we showed that when the pp-wave spacetime admits
a proper CKV, then the conformal factor is a solution of the original
equation, which means that the proper CKVs, when there exist, generate
always Lie and Noether point symmetries for the wave equation (\ref{wave.01}%
).

The Lie and Noether point symmetries of the wave equation (except the
trivial ones) for the isometry classes 1 to 14, of section \ref%
{isometryclass} are given in table \ref{WaveSym1}.

\begin{table}[tbp] \centering%
\caption{Lie and Noether point symmetries for the wave equation in the
pp-wave spacetime (\ref{pp.01}), for the isometry classes of \cite{ppKV}}%
\begin{tabular}{|c|c|c|c|c|c|}
\hline\hline
\textbf{Class} & $\mathbf{\#}$\textbf{\ } & \textbf{Lie/Noether Sym.} &
\textbf{Class} & $\#$ & \textbf{Lie/Noether Sym.} \\ \hline
\textbf{1} & $1$ & $k$ & \textbf{6iv} & $4$ & $k,~X_{2}^{\left( 6\right)
},~X_{3}^{\left( 6\right) },~Y_{4}^{\left( 6iv\right) }$ \\
\textbf{1i} & $2$ & $k,~X_{2}^{\left( 1i\right) }$ & \textbf{7} & $3$ & $%
k,~X_{2}^{\left( 7\right) },~X_{3}^{\left( 7\right) }$ \\
\textbf{2} & $2$ & $k,~X_{2}^{\left( 2\right) }$ & \textbf{8} & $3$ & $%
k,~X_{2}^{\left( 8\right) },~X_{3}^{\left( 8\right) }$ \\
\textbf{2i} & $3$ & $k,~X_{2}^{\left( 2\right) },~Y_{3}^{\left( 2i\right)
}/Y_{3\left( -1\right) }^{\left( 2i\right) }$ & \textbf{8}$_{\left( 0\right)
}$ & $4$ & $k,~X_{2}^{\left( 8\right) },~X_{3}^{\left( 8\right)
},~X_{4}^{\left( 8\right) }$ \\
\textbf{2ii} & $3$ & $k,~X_{2}^{\left( 2\right) },~Y_{3}^{\left( 2ii\right)
}/Y_{3\left( 0\right) }^{\left( 2ii\right) }$ & \textbf{8i} & $5$ & $%
k,~X_{2}^{\left( 8\right) },~X_{3}^{\left( 8\right) },~X_{4}^{\left(
8\right) },~Y_{5}^{\left( 8i\right) }$ \\
\textbf{2iii} & $3$ & $k,~X_{2}^{\left( 2\right) },~Y_{3}^{\left(
2iii\right) }$ & \textbf{8ii} & $6$ & $k,~X_{2}^{\left( 8\right)
},~X_{3}^{\left( 8\right) },~X_{4}^{\left( 8\right) },~X_{5}^{\left(
8ii\right) },~X_{6}^{\left( 8ii\right) },Y_{5}^{\left( 8i\right) }~$ \\
\textbf{3} & $2$ & $k,~X_{2}^{\left( 3\right) }$ & \textbf{8iii} & $6$ & $%
k,~X_{2}^{\left( 8\right) },~X_{3}^{\left( 8\right) },~X_{4}^{\left(
8\right) },~Y_{5}^{\left( 8i\right) },~Y_{6}^{\left( 8iii\right) }$ \\
\textbf{4} & $2$ & $k,~X_{2}^{\left( 4\right) }$ & \textbf{9} & $5$ & $%
k,~X_{2}^{\left( 9\right) },~X_{3}^{\left( 9\right) },~X_{4}^{\left(
9\right) },~X_{5}^{\left( 9\right) }$ \\
\textbf{5} & $3$ & $k,~X_{2}^{\left( 5\right) },~X_{3}^{\left( 5\right) }$ &
\textbf{10} & $6$ & $k,~X_{a}^{\left( 10\right) },~Y_{6}^{\left( 10\right) }$
\\
\textbf{5i} & $4$ & $k,~X_{2}^{\left( 5\right) },~X_{3}^{\left( 5\right)
},~Y_{4}^{\left( 5i\right) }$ & \textbf{10i} & $7$ & $k,~X_{a}^{\left(
10\right) },~H_{6}^{\left( 10\right) },~Y_{7}^{\left( 10i\right) }$ \\
\textbf{5ii} & $4$ & $k,~X_{2}^{\left( 5\right) },~X_{3}^{\left( 5\right)
},~Y_{4}^{\left( 5ii\right) }$ & \textbf{10ii} & $7$ & $k,~X_{a}^{\left(
10\right) },~H_{6}^{\left( 10\right) },~Y_{7}^{\left( 10ii\right) }$ \\
\textbf{6} & $3$ & $k,~X_{2}^{\left( 6\right) },~X_{3}^{\left( 6\right) }$ &
\textbf{11} & $7$ & $k,~X_{a}^{\left( 10\right) },~H_{6}^{\left( 10\right)
},~X_{7}^{\left( 11\right) }$ \\
\textbf{6i} & $5$ & $k,~X_{2}^{\left( 6\right) },~X_{3}^{\left( 6\right)
},~Y_{4}^{\left( 6i\right) },~Y_{5}^{\left( 6i\right) }$ & \textbf{12} & $7$
& $k,~X_{a}^{\left( 10\right) },~H_{6}^{\left( 10\right) },~X_{7}^{\left(
12\right) }$ \\
\textbf{6ii} & $5$ & $k,~X_{2}^{\left( 6\right) },~X_{3}^{\left( 6\right)
},~Y_{4}^{\left( 6ii\right) },~Y_{5}^{\left( 6ii\right) }$ & \textbf{13} & $%
7 $ & $k,~X_{a}^{\left( 10\right) },~H_{6}^{\left( 10\right)
},~X_{7}^{\left( 13\right) }$ \\
\textbf{6iii} & $4$ & $k,~X_{2}^{\left( 6\right) },~X_{3}^{\left( 6\right)
},~Y_{4}^{\left( 6iii\right) }$ & \textbf{14} & $7$ & $k,~X_{a}^{\left(
10\right) },~H_{6}^{\left( 10\right) },~X_{7}^{\left( 14\right) }$ \\
\hline\hline
\end{tabular}%
\label{WaveSym1}%
\end{table}%

\section{Conclusions}

\label{conclusion}

In this work we performed a complete classification of the Lie /Noether
point symmetries for the Klein-Gordon and the wave equation in pp-wave
spacetimes using three results: (a) The general results of \cite{IJGMMP} and
\cite{IJGMMP2} concerning the relation between the Lie / Noether point
symmetries of the Klein-Gordon equation with the conformal algebra of the
underlying space; (b) the classification of the Klein Gordon equation based
on the isometries of (\ref{pp.01}) done in \cite{ppKV}, and (c) The
classification of the conformal algebra of the pp-wave spacetimes (\ref%
{pp.01}) done in \cite{ppCKV} and \cite{ppnull}.

In addition we used these results in order to calculate the Lie and the
Noether point symmetries of the wave equation (\ref{wave.01}) in a pp-wave
spacetime. \ We found that the Lie point symmetries form a Lie algebra $%
G_{W}~$(except the trivial symmetries), of dimension $\dim G_{W}\leq 7$
where the equality holds for the case where the space (\ref{pp.01}) is a
plane wave spacetime. In addition we noted that due to the fact that the
conformal factors of the CKVs of (\ref{pp.01}) are solutions of wave
equation (\ref{wave.01}) all CKVs of (\ref{pp.01}) give rise to a Lie point
symmetry give a Lie point symmetry of (\ref{wave.01}). Because we have
followed the classification of \cite{ppCKV}, the symmetry classification
holds and for non-empty spacetimes.\

A further use of the results obtained in this work is that they can be used
in order one to reduce and possibly to solve analytically the Klein-Gordon
equation (\ref{pp.02}) and wave equation (\ref{wave.01}) in a pp-wave
spacetime.

{\large {\textbf{Acknowledgements}}} \newline
The research of AP was supported by FONDECYT postdoctoral grant no. 3160121.

\end{document}